\documentclass[aps,twocolumn,prb,superscript,floatfix,superscriptaddress,showpacs,footinbib]{revtex4-2}

\usepackage{amssymb}

\usepackage{graphicx}
\usepackage{amsmath}
\usepackage{amsmath,bm}
\usepackage{blindtext, xcolor}
\usepackage{xcolor}
\usepackage{hyperref}
\hypersetup{
	colorlinks,%
	citecolor=blue,%
	linkcolor=blue,%
	urlcolor=blue
}

\begin{document}
	
%
\title{Majorana correlations in quantum impurities coupled to a topological wire}%
	
\author{G. S. Diniz}%
\email[e-mail:]{ginetom@gmail.com}
\affiliation{Curso de F\'isica, Universidade Federal de Jata\'i, Jata\'i, GO 75801-615, Brazil}
\affiliation{Instituto de F\'isica,  Universidade Federal  de Uberl\^andia,Uberl\^andia, MG 38400-902, Brazil}
\author{E. Vernek}%
\email[e-mail:]{vernek@ufu.br}
\affiliation{Instituto de F\'isica,  Universidade Federal  de Uberl\^andia,Uberl\^andia, MG 38400-902, Brazil}
\affiliation{Department of Physics and Astronomy, and Nanoscale and Quantum Phenomena Institute, Ohio University, Athens, Ohio 45701-2979, USA}
\date{\today}%

\begin{abstract}
In this work we investigate Majorana correlations in quantum impurities coupled to a topological superconducting wire. By employing a density matrix renormalization group approach, we have calculated the Majorana correlation functions in the ground state of the system for different positions of the impurities along the wire. We observe that the correlations decrease exponentially with the distance between the impurities and the ends of the wire. Moreover, different electron-electron interactions on the wire and on the impurities were also analyzed, and the results showed a stronger effect as the electron occupation in the wire is increased. In addition, while changing the occupation of the impurities by tuning their energy levels, we observe that for specific values of $\varepsilon$ the absolute value of correlation is highly peaked when the chemical potential lies within the window of the topological regime. These peaks are associated to Coulomb blockade phenomena in the impurities, revealing very clearly its effect on the Majorana bound states in the topological phase of the system.

\end{abstract}

\maketitle

\section{Introduction}
Since the seminal work by Kitaev~\cite{Kitaev_2001}, which predicted the emergence of Majorana bound states (MBSs) at the \emph{ends} of a $p$-wave superconductor spinless chain, an enormous effort has been devoted to endeavor realistic platforms and experimental techniques to probe such exquisite quasi-particle excitations~\cite{RevModPhys.80.1083,Sato2017,Prada2020,Breunig2022}. Most of the intense interest relies on the 
outrageous physical properties displayed by the Majorana fermions (MFs), for instance, a particle being its own antiparticle and the emergent non-Abelian statistics~\cite{10.1063/5.0055997}. These properties have attracted a great deal of attention in the recent years, specially because of their potential application in topological quantum computation~\cite{KITAEV20032,RevModPhys.80.1083}.

In spite of the great number of theoretical and experimental investigations carried on in the past decade, detection and manipulation of MBSs is a challenging experimental task. This is because the smoking gun signature of MBSs --- a zero energy conductance peak in transport experiment --- can be associated to different phenomena, thus poisoning a unambiguous detection of  MBS~\cite{science.aax6361,PhysRevLett.126.076802,science.abf1513}. To circumvent these difficulties, different theoretical and experimental approaches have been proposed to detect MBSs in different physical systems~\cite{Prada2020}, for instance, tunneling spectroscopy~\cite{Gul2018,Grivnin2019}, cross-correlation shot noise~\cite{PhysRevLett.124.096801} and other transport related experiments~\cite{Prada2020}.

Among the various physical platforms proposed to show the emergence of Majorana bound states (MBS), one dimensional (1D) semiconductor nanostructures has shown to be very attractive on both theoretical and experimental investigations~\cite{PhysRevLett.105.177002,PhysRevB.96.195430,10.1126/science.1222360,10.1126/science.aaf3961}. To realize a topological superconductor quantum wire, a semiconductor nanowire with strong Rashba spin-orbit interaction (SOC) is brought into proximity of a $s$-wave superconductor. Thanks to SOC, the $s$-wave superconductor thus induces $p$-wave pairing in the wire. Then, by appropriate application of an external magnetic field, the system enters into a topological phase, within which MBSs emerge~\cite{PhysRevLett.105.077001}. It has also been shown in such a system that Majorana modes leak into quantum dots (QDs) attached to the topological superconductor  wires~\cite{PhysRevB.89.165314,PhysRevB.91.115435} which could be experimentally accessed through transport experiments. This phenomena, that is very robust against local $e$-$e$ interactions and gate potential, provides a suitable manner to detect MBS in quantum wires~\cite{PhysRevB.91.115435,PhysRevB.101.075428}.

Recently, Wang \emph{et al.}~\cite{PhysRevB.96.205428} have theoretically demonstrated how to characterize distinct phases (trivial, Su-Schrieffer-Heeger-like topological and topological superconductor phases) in a dimerized Kitaev model by inspecting correlation functions (for the two edges) of both the noninteracting and interacting regimes. Interestingly, several studies on quantum correlations phenomena on MBSs emergent systems have demonstrated a rich physics that suggests alternative manner to probe such exquisite states~\cite{PhysRevLett.111.056802,PhysRevB.91.214507,PhysRevB.96.085117,doi:10.7566/JPSJ.86.124715,Reslen_2018,Miao2018}. For instance, an all-electric manipulation of \emph{nonlocal} spin correlations has been demonstrated between the ends of two quantum wires, when in close proximity to a 2D unconventional superconductor~\cite{PhysRevLett.110.117002}. In addition, current-correlations measurements have been proposed as a suitable probe to distinguish MBS from Andreev bound states~\cite{PhysRevB.96.115413}, as negative current-correlation would be associated to exotic states. Also, exploring the \emph{nonlocal} nature of MFs, quantum discord of two QDs mediated by the paired Majorana fermions have been proposed as a measurable physical quantity to probe the presence of MFs~\cite{Li2014}.

Motivated by the variety of physical scenarios that can be captured by quantum correlations, in this work we aim at showing how we can explore Majorana correlations to detect the emergence of MBS in quantum impurities attached to a topological wire. We propose a system composed of two impurities coupled to a quantum wire that can be driven into the topological superconducting phase.
We calculate the spin-resolved Majorana correlation functions between the impurities in both interacting and noninteracting regimes of the system. The results show a clear signature of the MBSs leaked into the impurities, which is translated into sharp peaks in the correlations and strongly enhanced as the impurities are coupled closer to the edge of the topological wire. These signatures are robust against $e$-$e$ interactions in the low electron occupation regime in the impurities. More interestingly, we observe that correlation reveals a very interesting physics related to  coulomb blockade phenomena in the MBS leaked into the impurities.

The remaining of this paper is organized as follows.  In Sec.~\ref{model} we introduce our theoretical model and methods, in Sec.~\ref{results} we present and discuss our numerical results. Finally, in Sec.~\ref{conclusions} we bring our concluding remarks.

\section{Theoretical Model}
\label{model}
\begin{figure}[h!]
\includegraphics[scale=0.55]{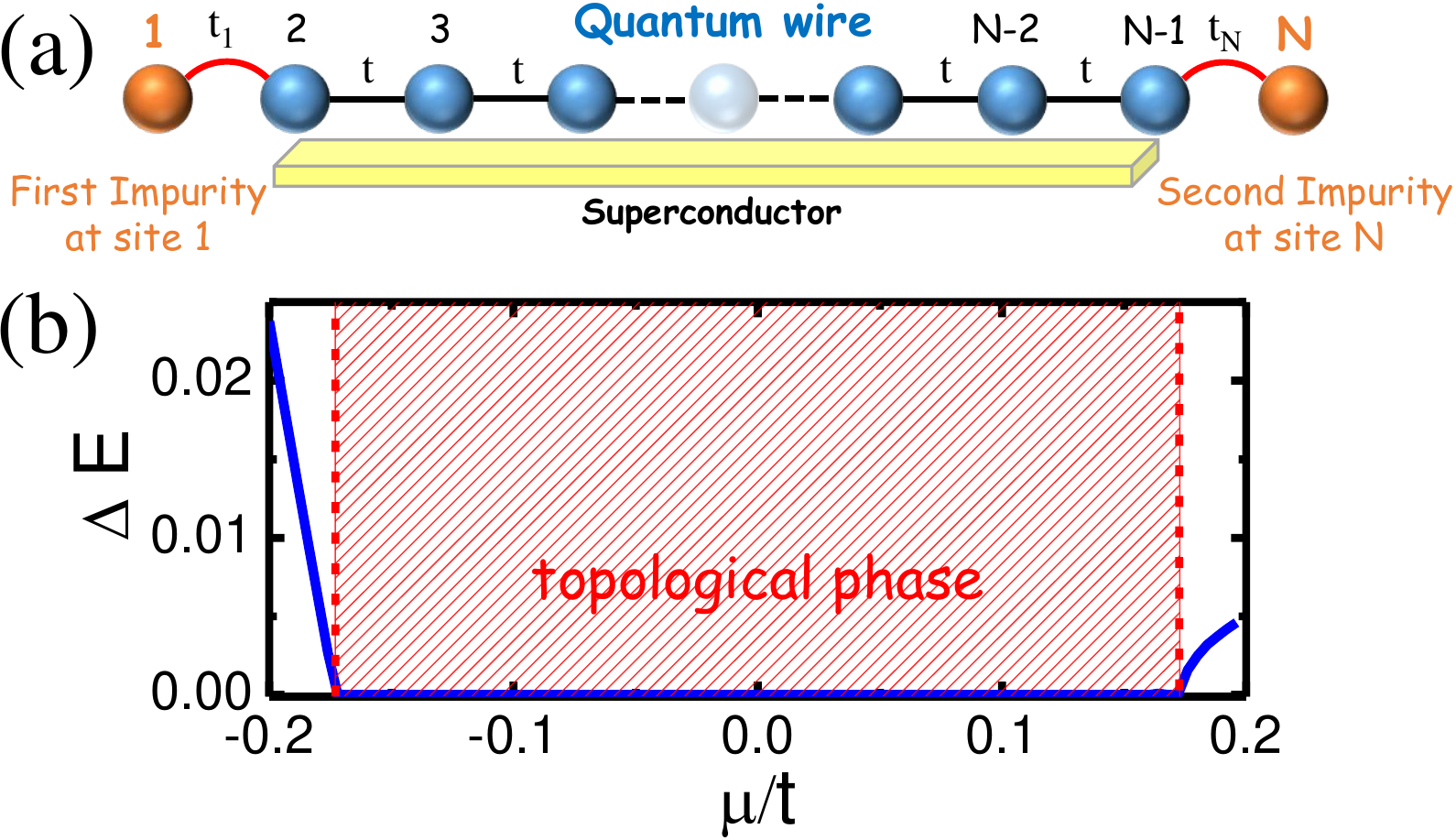} 
\caption{(a) Schematic representation of two impurities (at site $1$ and site $N$) coupled to a quantum wire (chain) in close proximity to a superconductor. (b) The panel shows the $\Delta E$ for the chain (N=160), as function of $\mu$ for the non-interacting chain (without any impurity attached). For the chemical potential inside the window (vertical dashed red line) the system is in the topological phase. The parameters used for the lower panel are: $t=1.0$, $\Delta$=0.1t, $\lambda_{\rm SO}$=0.1t, $V_{Z}$=0.2t. 
\label{fig0}}
\end{figure}

The system is comprised of a one dimensional semiconductor wire with Rashba spin-orbit interaction. The system is placed in close proximity to a $s$-wave superconductor, hence an electron pairing potential $\Delta$ is induced in the wire \cite{Lutchyn2018}. This system is described by a tight-binding model consisting of a total of $N-2$ atomic sites.  The two impurities (which we label as sites $i=1$ and $i=N$) is attached to the quantum wire (in the lattice model) directly to any site of the chain, from $i=2$ to $i=N-1$, as shown in  Fig.~\ref{fig0}(a). The full Hamiltonian of the system can be written as,
%
%
\begin{eqnarray}
H = H_{\rm C} + H_{\rm I} + H_{\rm CI},
\label{eq:Hamilt0}
\end{eqnarray}
where $H_{\rm C}$, $H_{\rm I}$ and $H_{\rm CI}$ stand for the topological quantum wire (chain), the impurities and chain-impurity coupling, respectively.  Explicitly, in the Hamiltonian of Eq.~\eqref{eq:Hamilt0}, 
$H_{\rm C}$ is given by~\cite{PhysRevB.84.014503},
\begin{eqnarray}
H_{\rm C}&=&-\sum_{i\sigma}(\mu - t)c^\dag_{i\sigma}c_{i\sigma} -\frac{t}{2}\sum_{\langle i,j\rangle,\sigma}c^\dag_{i\sigma}c_{j\sigma}\\ \nonumber  
 &+&V_{\rm Z}\sum_{i\sigma\sigma^{\prime}}\sigma^{z}_{\sigma \sigma^\prime}c^\dag_{i\sigma}c_{i\sigma^{\prime}} + \frac{i\lambda_{\rm SO}}{2}\sum_{\langle i,j\rangle, \sigma}c_{i\sigma}^{\dagger }\sigma^{y}_{\sigma\sigma^\prime}c_{j\sigma}\\ \nonumber
&+& \Delta\sum_{i}\left(c_{i\uparrow}^{\dagger}c_{i\downarrow}^{\dagger} + H.c.\right)+ U\sum_{i}n_{i\uparrow}n_{i\downarrow}.
\label{eq:Hamilt1}
\end{eqnarray}
Here, $c_{i\sigma}^{\dagger}$ ($c_{i\sigma}$) creates (annihilates) an electron with spin $\sigma$ at site $i$, $\mu$ is the chemical potential in the quantum chain, $t$ is the hopping matrix element between nearest neighbors,  $V_{\rm Z}$ is the Zeeman energy,  $\lambda_{\rm SO}$ is the strength of the Rashba spin-orbit interaction,  $\Delta$ is the superconductor pairing potential and $U$ accounts  for the on-site $e$-$e$ interaction in the quantum chain. Finally, $\sigma^y$ and $\sigma^z$ are the known Pauli matrices. The second term of the Eq.~\eqref{eq:Hamilt0} is given by,

\begin{eqnarray}
H_{\rm I}&=& \sum_{i=\{1,N\},\sigma}(\varepsilon_{i} +t) c^\dag_{i\sigma}c_{i\sigma} + U_{\rm I}\sum_{i=\{1,N\},\sigma}n_{i\uparrow}n_{i\downarrow}, 
\label{eq:Hamilt2}
\end{eqnarray}

where, $\varepsilon_{i}$ is the impurities' on-site energy, while and $U_{\rm I}$ stands for the $e$-$e$ interaction in the impurities, which is assumed to be equal for both impurities, and $n_{i\sigma}=c^\dagger_{i\sigma}c_{i\sigma}$ is the number operator at site $i$.

Finally, for the third therm of Eq.~\eqref{eq:Hamilt0} has the form
\begin{eqnarray}
H_{\rm CI}\!=\!-\sum_{\sigma}(t_{1}c^\dag_{1\sigma}c_{j\sigma} + t_{N}c^\dag_{N\sigma}c_{j^{\prime}\sigma}). 
\label{eq:Hamilt3}
\end{eqnarray}
Here, $t_{1}$ and $t_{N}$  represent the hopping of an electron between the impurities and the chain. Moreover, in the above $ j,j^\prime \in [2,N-1]$ where the site $2$ and $N-1$ correspond to the ends of the chain.

While the full Hamiltonian can be solved in an exact manner in the non-interacting case $(U=U_1=U_N=0)$, in the presence interacting terms this is no longer possible. Here, we are interested in the low-energy regime of the system, thus, obtaining the many-body ground state (in some cases also the few excited states) is good enough for our purpose. This can be achieved very accurately within Density Matrix Renormalization Group (DMRG) calculations. Our  DMRG results are obtained with the numerical package {ITensor}~\cite{itensor,note1}, written within  matrix product states (MPS) formalism \cite{PhysRevLett.75.3537}. Having the many-body ground state, we then calculate the Majorana correlation matrix for the entire system. In particular, we focus on the correlations between MF located at the ends of the topological wire, as well as the correlation between MFs leaked into the impurities. 
Let us define the MF operators~\cite{Silva_2016}
\begin{eqnarray}
\gamma_{A,i}^{\sigma}=\frac{1}{2}\left(c_{i\sigma} + c_{i\sigma}^{\dagger}\right) \,\,\, \mbox{and} \,\,\,
\gamma_{B,i}^{\sigma}=-\frac{i}{2}\left(c_{i\sigma} - c_{i\sigma}^{\dagger}\right),
\label{eq:majop}
\end{eqnarray}
in which $\gamma_{\alpha,i}^{\sigma}=(\gamma_{\alpha,i}^{\sigma})^{\dagger}$ (with $\alpha=A,B$) are Majorana operators obeying the anti-commutation relation $[\gamma_{\alpha,i}^{\sigma},\gamma_{\alpha^{\prime},j}^{\sigma^{\prime}}]_+=2\delta_{ij}\delta_{\alpha\alpha^{\prime}}\delta_{\sigma\sigma^{\prime}}$. With this operators we define the single-particle  MF correlation functions,
%
%
%
\begin{eqnarray}
C^{\sigma\sigma^{\prime}}_{\alpha i;\beta j}\equiv \langle\gamma_{\alpha,i}^{\sigma}\gamma_{\beta,j}^{\sigma^{\prime}}\rangle,
\end{eqnarray}
in which $\sigma,\sigma^{\prime}=\uparrow,\downarrow$, $i (j)$ is the site in the chain and  $\alpha,\beta \in \{A,B\}$. In the equation above, $\langle \cdots \rangle\equiv \langle \Psi_0|\cdots |\Psi_0\rangle$, which $ |\Psi_0\rangle$ represents the many-body ground states of the  system, calculated within DMRG.


%

\section{Results and Discussion}
\label{results}
To obtain our numerical results, we set the hopping matrix element $t=1$ as our energy unit and, following Ref.~\cite{PhysRevB.84.014503}, we also set $\lambda_{\rm SO}$=0.1t, $V_{Z}=0.2$ and $\Delta=0.1$. For simplicity, we will also consider $t_1=t_N\equiv t_0$ and $\varepsilon_{1}=\varepsilon_{N}\equiv\varepsilon$. The trivial-to-topological quantum phase transition will be driven by a suitable control of $V_Z$, $\Delta$ and $\mu$.  
The results are for a finite chain (quantum wire) with $N$ number of sites, where the first and last sites stand for the impurities. 

{\it Non-interacting case ---}
For the non-interacting case ($U=U_{I}$=0), the topological superconducting phase with the emergence of MBS in the chain is accessed when the parameters of the chain $\Delta$, $\mu$, $V_{Z}$ fulfill the condition $-\mu_c < \mu < \mu_c$, where   $\mu_c=\sqrt{V_{Z}^{2}-\Delta^{2}}$~\cite{PhysRevLett.105.077001,PhysRevLett.105.177002}. For the sake of completeness, we first check the region in the parameter space in which  the system is in the topological regime. According to Ref.~\cite{PhysRevB.84.014503}, there are distinct ways to identify the superconducting topological state in the chain within DMRG approach. Here, we use the gap $\Delta E=E^{\rm even}_0-E^{\rm odd}_0$, where $E^{\rm even}_0$ and $E^{\rm odd}_0$ are the many-body lowest energies within the Hilbert subspaces with even and odd number of electrons, respectively.  Since the total Hamiltonian does not mix these two subspaces, we can access these energies by starting the DMRG variational procedure with an state within each to these sectors of the Hilbert space. In the topological phase, the ground state is expected to be doubly degenerated, since the two Majorana zero modes bound to the edges of the quantum wire compose a conventional doubly degenerate Fock space $\{|0\rangle,|1\rangle \}$, thus rendering $\Delta E=0$.

\begin{figure}[h!]
\includegraphics[scale=0.50]{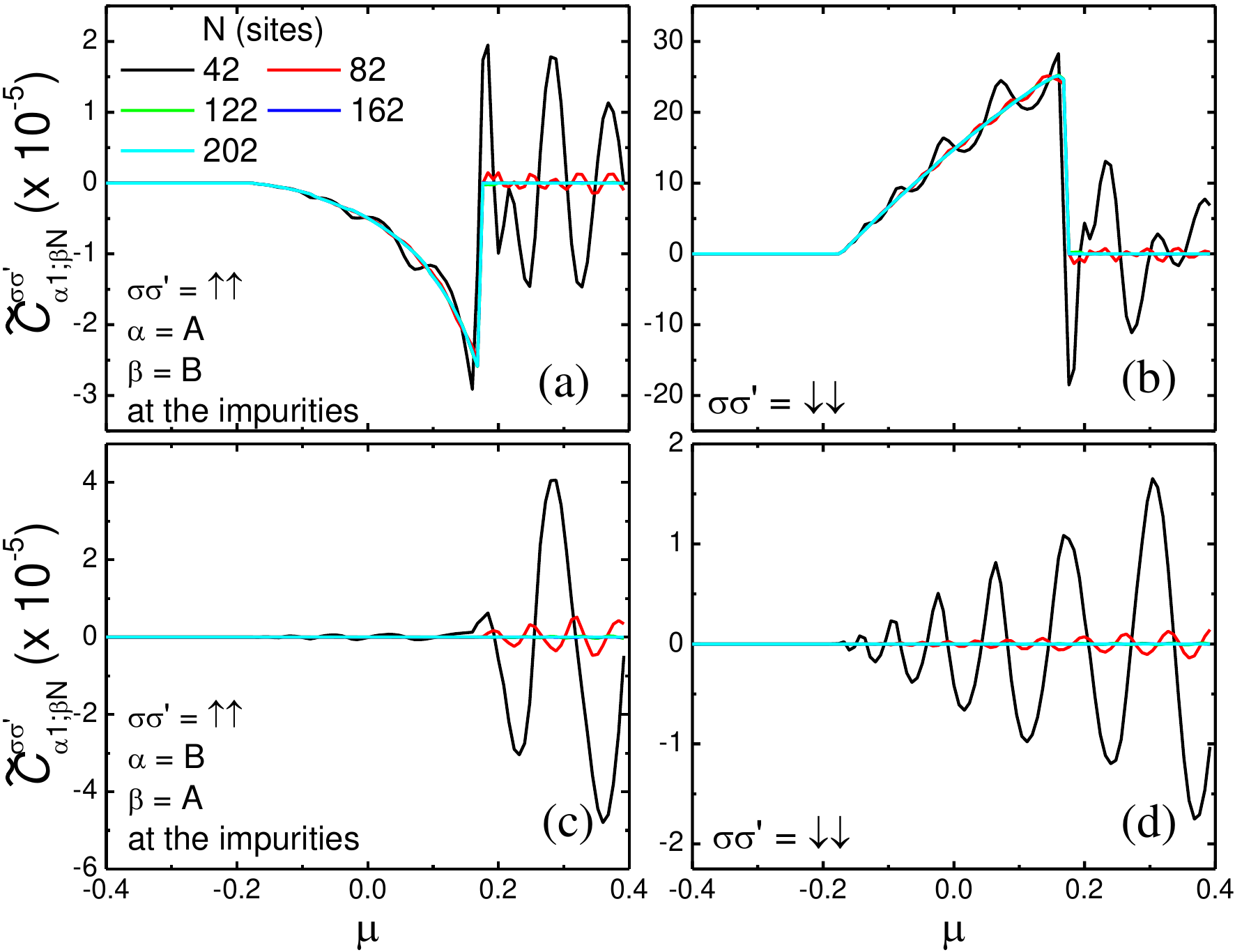}
\caption{Imaginary part of the Majorana correlations as function of chemical potential $\mu$ for different number of sites and $\alpha$ ($\beta$) components. (a) for $\tilde{C}_{Ai;Bj}^{\uparrow\uparrow}$, (b) for  $\tilde{C}_{Ai;Bj}^{\downarrow\downarrow}$, (c) for $\tilde{C}_{Bi,Aj}^{\uparrow\uparrow}$ and (d) for  $\tilde{C}_{Bi;Aj}^{\downarrow\downarrow}$. The parameters used are: $\Delta$=0.1, $\lambda_{\rm SO}$=0.1t, $V_{Z}=0.2$, $\varepsilon=0$ and $t_{0}=0.1$.} 
\label{fig1}
\end{figure}
%
In Fig.~\ref{fig0}(b) we show  $\Delta E$ as function of the chemical potential $\mu$ for $U=0$ for a chain of 160 sites (without impurities). Here we have decoupled the impurities  from the chain by setting $t_0=0$. Note that only within the region of of $-0.173 \lesssim \mu \lesssim 0.173$, $\Delta E=0$, indicating the topological phase of the system.

Once the topological phase of the quantum wire is confirmed, we now attach the impurities and calculate the MFs correlation function $C^{\sigma\sigma^{\prime}}_{\alpha 1,\beta N}$ for different parameters configurations.  It is known that in short wires the Majorana edges modes may couple together and splitting off the zero-energy degeneracy. 

{To verify how the length of the chain impacts on the MF correlations, in Fig.~\ref{fig1} we show the  $C^{\sigma\sigma^{\prime}}_{\alpha 1;\beta N}$ with spin components $\sigma\sigma^{\prime}=\uparrow\uparrow,\downarrow\downarrow$, as function of chemical potential $\mu$ for different values of $N$. By direct calculations, we have checked that any finite MF correlation is purely imaginary. Therefore, in the following we show only their imaginary parts $\tilde C^{\sigma\sigma^{\prime}}_{\alpha 1;\beta N}={\rm Im}(C^{\sigma\sigma^{\prime}}_{\alpha 1;\beta N})$. Figures~\ref{fig1}(a) and  ~\ref{fig1}(b) we show $\tilde C^{\uparrow\uparrow}_{A1;BN}$ and $\tilde  C^{\downarrow\downarrow}_{A1;BN}$, respectively. Similarly, Figs.~\ref{fig1}(c) and \ref{fig1}(d) shows, respectively, $\tilde C^{\uparrow\uparrow}_{B1;AN}$ and $\tilde C^{\downarrow\downarrow}_{B1;AN}$. We first note that  all MF correlations vanish for $\mu < -\mu_c$. Inside the topological phase ($-\mu_c < \mu < \mu_c$) there is an increase (in absolute value) of the correlations $\tilde C^{\uparrow\uparrow}_{A1;BN}$ and $\tilde C^{\downarrow\downarrow}_{A1;BN}$. Furthermore, note that $\tilde C^{\downarrow\downarrow}_{A1;BN}$ is about one order of magnitude larger that $\tilde C^{\uparrow\uparrow}_{A1;BN}$ for all values of $N$ [except for the panels \ref{fig1}(c) and \ref{fig1}(d)]. This is because for $V_Z$ positive, the chain is polarized negatively. For $\mu> \mu_c$ we note an abrupt decrease of the MF correlations, vanishing for large $N$ (see for instance the case of $N=202$, shown in cyan line). We also observe quite strong oscillations for small N [see $N=42$ (black line) for instance]. These oscillations disappear completely for larger $N$, signaling an important finite-size effect, (specially because the overlap of the MBSs is exponentially suppressed for longer chains).  Since sites $1$ and $N$ correspond to the impurities, the appearance of the MF correlations results from the leaking of the Majorana modes from the end of the chain into the impurities. Now, observe in Figs.~\ref{fig1}(c) and \ref{fig1}(d) that $\tilde C^{\uparrow\uparrow}_{B1;AN}$ and $\tilde C^{\downarrow\downarrow}_{B1;AN}$ indicate nothing special in the topological phase. Indeed they vanish within the entire interval of $\mu$ for larger $N$. Similar to the previous results, oscillations are also observed for small $N$. We should also mention that, consistently with the Kitaev model, by changing  $\Delta \rightarrow -\Delta$, $\tilde C^{\sigma\sigma^{\prime}}_{A1;BN}$ vanish while $\tilde C^{\sigma\sigma^{\prime}}_{B1;AN}$ becomes finite \cite{Kitaev_2001}.}
\begin{figure}[h!]
\includegraphics[scale=0.50]{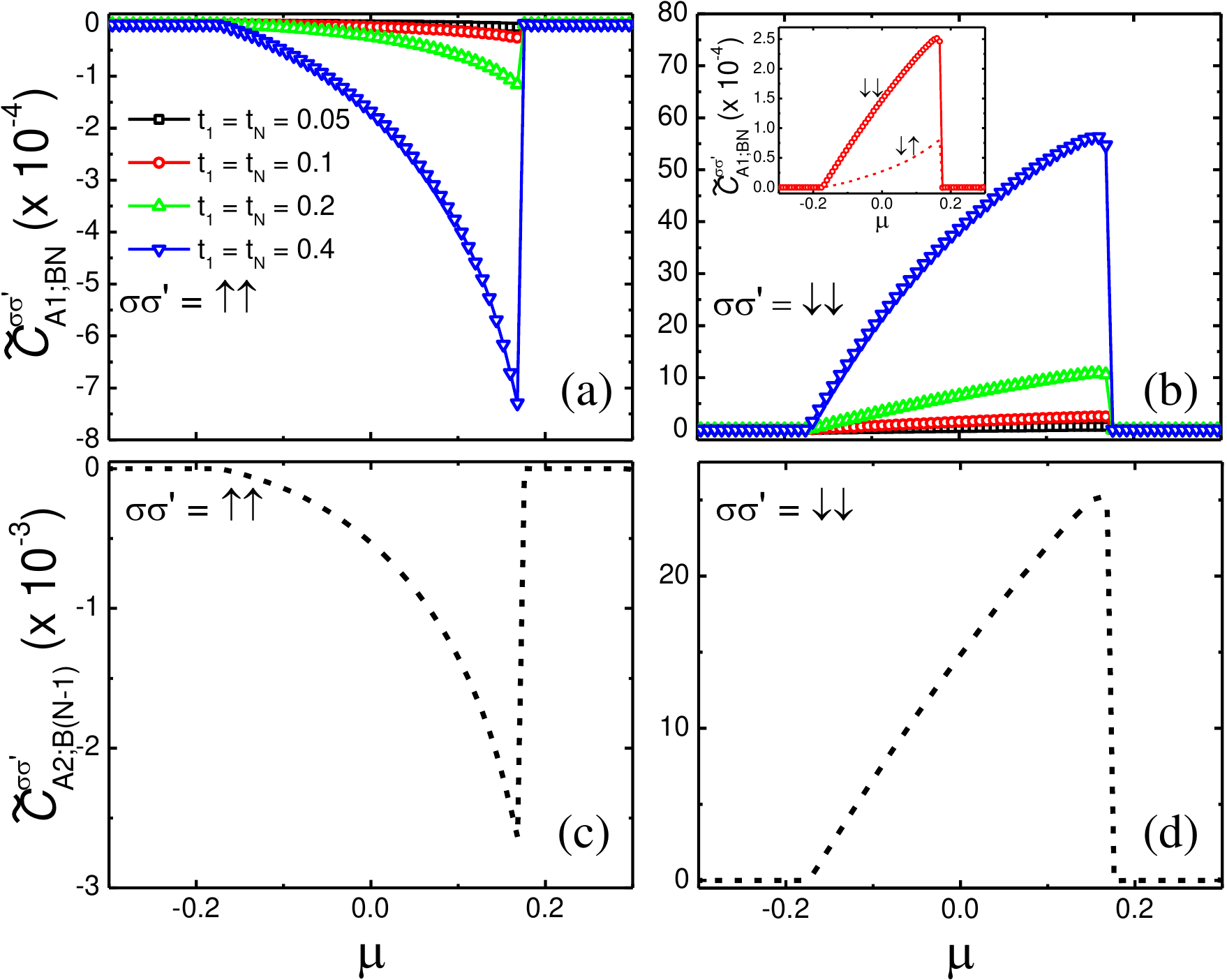}
\caption{Imaginary part of the Majorana correlations as function of chemical potential $\mu$ for different impurity-chain coupling $t_{i}$, where $t_{1}=t_{N}=t_{0}$ is assumed. (a) $\tilde{C}_{A1;BN}^{\uparrow\uparrow}$ and (b)  $\tilde{C}_{A1;BN}^{\downarrow\downarrow}$. (c) and (d) are the edge-edge Majorana correlation, i.e. $\tilde{C}_{A2;B161}^{\sigma\sigma^{\prime}}$, for $\sigma\sigma^{\prime}$= $\uparrow\uparrow$ and $\downarrow\downarrow$ with $t_0=0.1$, respectively. The inset in panel (b) shows a zoom of the curve with $t_{1}=t_{N}=0.1$ and with spin components $\downarrow\downarrow$ and $\downarrow\uparrow$. The parameters used are: $N=162$, $\Delta=0.1$, $\lambda_{\rm SO}=0.1$, $V_{Z}=0.2$, $\varepsilon=0$.} 
\label{fig2}
\end{figure}
%


From what we have seen in Fig.~\ref{fig1}, for a long enough chain there is no substantial change in the MF correlations (compare for instance the correlations for $N=122$ and $N=202$). Therefore, we will now set the number of sites in the chain to $N=162$ (unless stated otherwise). To inspect how the MF correlations change as strength of  the impurity-chain coupling changes, in Figs.~\ref{fig2}(a) and \ref{fig2}(b) we show $\tilde C^{\uparrow\uparrow}_{A1;BN}$ and  $\tilde C^{\downarrow\downarrow}_{A1;BN}$, respectively. Clearly, both correlations exhibit a very quickly enhancement as $t_0$ increases. In addition, as $t_{0}\rightarrow t$ one would expect that the leaked Majorana correlation becomes \emph{comparable} to the values of correlation at the chain's edge. In fact, this behavior can be noticed in Fig. \ref{fig2} (c)-(d), where is shown the correlation function between the two sites at the edges of the chain, for both $\sigma\sigma^{\prime}$= $\uparrow\uparrow$ and $\downarrow\downarrow$. This behavior is expected, as the impurity site would be {almost indistinguishable} (besides the SOC, Zeeman and superconductor terms at the chain) from any site of the chain, hence the leaking would be maximum. For increasing value of $t_0$, we see that the edge Majorana correlations further decreases while it increases between the two impurities. Again, the $\sigma\sigma^{\prime}$= $\downarrow\downarrow$ component shows to be one order higher as compared to the $\sigma\sigma^{\prime}$= $\uparrow\uparrow$ component. Importantly, in both correlations finite values are always constrained within the topological phase ($\mu < \vert\mu_c\vert$). Notice that since the impurities are coupled to the chain, which is in close proximity to a superconductor and with a strong Rashba SOC, there is also the possibility of an induced pairing correlation contribution and a \emph{opposite-spin} electronic correlation \cite{PhysRevB.78.035444,PhysRevB.106.155123}. Consequently, these \emph{opposite-spin} electronic correlations would be associated to a \emph{opposite-spin} MF correlation. To inspect these contributions, in the inset of Fig. \ref{fig2} (b) we show the curves for $\tilde{C}_{A1;BN}^{\downarrow\downarrow}$ and $\tilde{C}_{A1;BN}^{\downarrow\uparrow}$ (assuming the value $t_{1}=t_{N}$=0.1). As can be noticed, the \emph{opposite-spin} MF is three times smaller than the higher \emph{same-spin} contribution. This behavior could be observed for any finite value of $\lambda_{SO}$ and/or $\Delta$. In addition, the term $\tilde{C}_{A1;BN}^{\uparrow\downarrow}$ has the same strength, however with negative values. In the following results, we will focus on the \emph{same-spin} terms, because they will give quantitatively larger contributions, which is the case of $\tilde{C}_{\alpha i;\beta j}^{\downarrow\downarrow}$.
	

%
\begin{figure}[h!]
\includegraphics[scale=0.50]{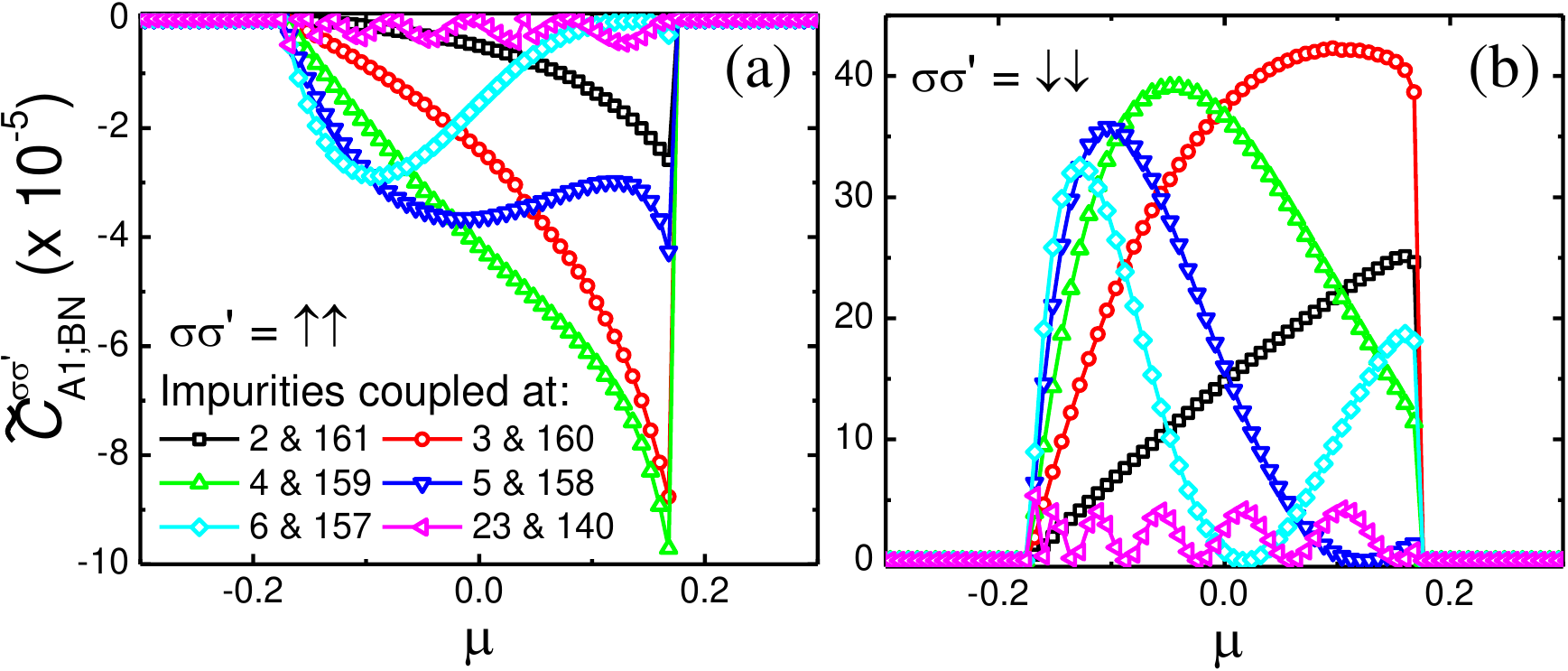}
\caption{Majorana correlations as function of chemical potential $\mu$, for different impurity-site coupling positions. (a) $C_{A1;BN}^{\uparrow\uparrow}$ and (b)  $C_{A1;BN}^{\downarrow\downarrow}$. The parameters used are: $N=162$, $\Delta=0.1$, $\lambda_{\rm SO}=0.1$, $V_{Z}=0.2$, $\varepsilon=0$ and $t_{0}=0.1$.} 
\label{fig3}
\end{figure}

The MF correlations studied before results from the leaking of the MBS from the end of the chain into the impurities. Therefore, the leaking depends on the position of the impurity relative to the ends of the chain. 
To verify this, in Figs.~\ref{fig3}(a) and \ref{fig3}(b) we show the MF correlations for different impurity locations. As the impurities moves to sites farther away from the edges (towards the center of the chain), the correlation  decrease and eventually vanishing for the inner sites. This occur because the localized wavefunction decays exponentially~\cite{PhysRevB.84.014503} as we increase the distance from the ends.

\textit{Effect of interactions ---} 
To address the effect of interaction in the Majorana correlations in th system we will three possible cases: (i) $e$-$e$ interaction only at the topological chain $U$, (ii) $e$-$e$ interaction only at the impurities sites with same value $U_{\rm I}$ for both impurities, and (III) with finite value of $e$-$e$ interaction in the impurity and the chain.
Let as first make $U$ finite in the chain. In this case,  it is expected that the pairing mechanism described by the term proportional to $\Delta$ in the Hamiltonian Eq. \eqref{eq:Hamilt1}  is suppressed (for $U>$0), reducing the effective bulk gap of the superconductor. As a result $\mu_c$ increases, rendering a wider range (along $\mu$-axis) of the topological phase for a fixed value of $V_Z$. At the same time, there is an increase in the wire magnetization~\cite{PhysRevLett.98.126408,PhysRevB.78.054436} that leads to  a more robust the topological phase against chemical potential fluctuations~\cite{PhysRevB.84.014503}. 
%
%
\begin{figure}[h!]
\includegraphics[scale=0.50]{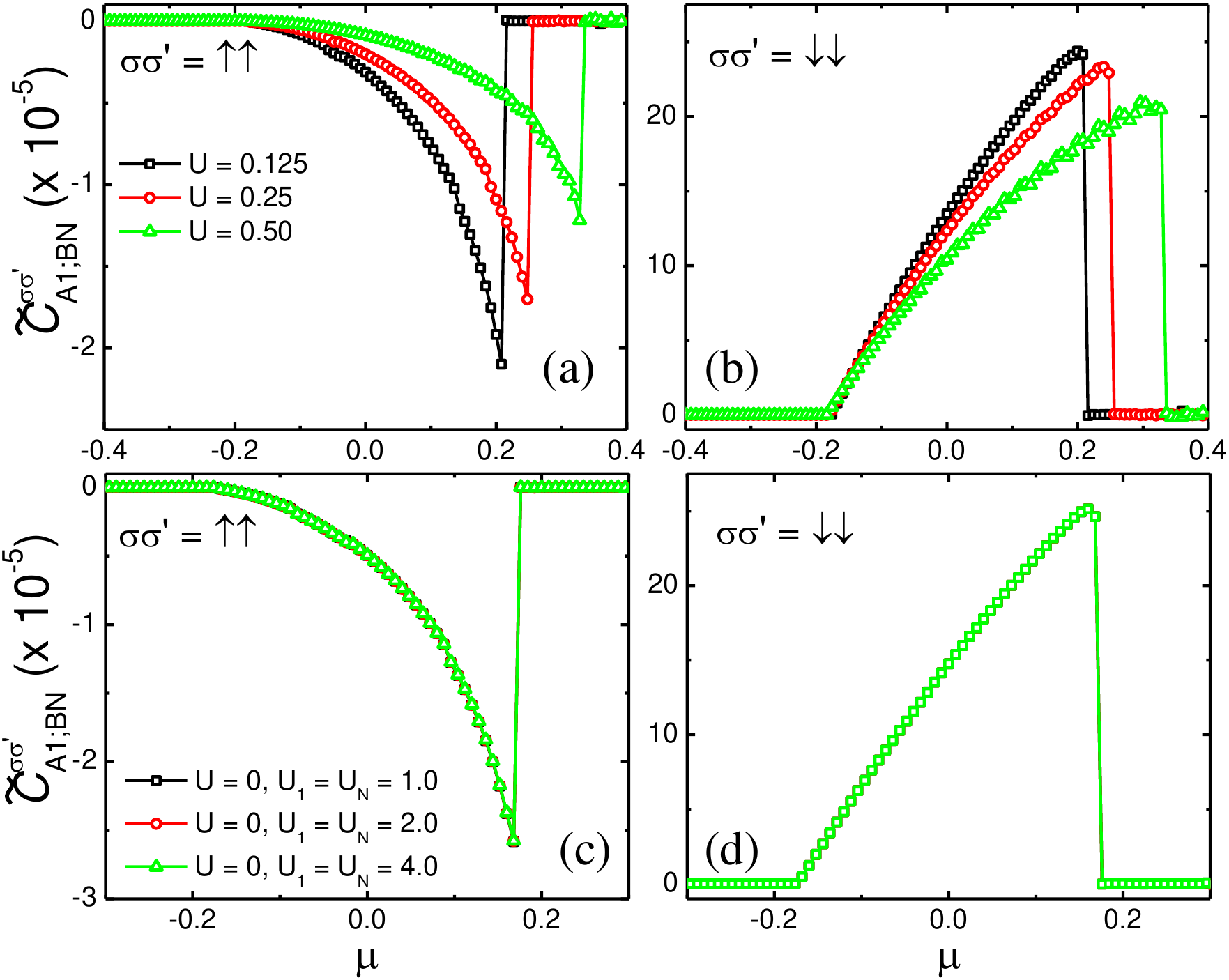}
\caption{(a) Majorana correlation at the impurity sites as function of chemical potential $\mu$, for different values of $U$ (with $U_{1}$=$U_{N}$=0) for spin $\sigma\sigma^{\prime}=\uparrow\uparrow$ and (b) for spin $\sigma\sigma^{\prime}=\downarrow\downarrow$. (c) and (d) are similar to the upper panels, but for different values of $U_{\rm I}$ (with $U$=0). The parameters used in all panels are: $N$=162, $\Delta=0.1$, $\lambda_{\rm SO}=0.1$, $V_{Z}=0.2$, $\varepsilon=0$ and $t_{0}=0.1$.} 
\label{fig4}
\end{figure}
%
%
The resulting effect is shown in Figs.~\ref{fig4}(a) and \ref{fig4}(b) which exhibit $\tilde{C}^{\uparrow\uparrow}_{A1;BN}$ and $\tilde{C}^{\downarrow\downarrow}_{A1,BN}$, respectively, for the impurity coupled to the sites at the end of the chain (sites 2 and 161) and for different values of $U$. As expected, as $U$ increases, we observe a widening in the range of the chemical potential where Majorana correlation is finite. Interestingly, note that only the upper boundary of the topological phase moves. According to Ref.~\cite{PhysRevB.84.014503}, this is because for  $\mu<$0 the electron occupation is very low (low electron density), therefore, Coulomb repulsion is not effective. On the other hand, $e$-$e$ interaction becomes very important in positive $\mu$, and increasing $U$ modifies the upper boundary of the topological phase substantially. 

Now, for the case where there is a repulsive interaction only at the impurities  ($U_{\rm I}>0$), we expect a smaller effect specially at low occupation ($\varepsilon >0$), as there are no electrons in the impurities to interact among each other. Truly, these can be noted in Fig.~\ref{fig4} (c)-(d), where the Majorana correlation for $\sigma\sigma^{\prime}$=$\uparrow\uparrow$ and  $\downarrow\downarrow$ components are shown to be almost identical. Notice that for these two panels, we have assumed a value of $U$=0, and $U_{I}$=$U_{1}=U_{N}$=0.0, 2.0 and 4.0. 

\begin{figure}[t!] 
\includegraphics[scale=0.55]{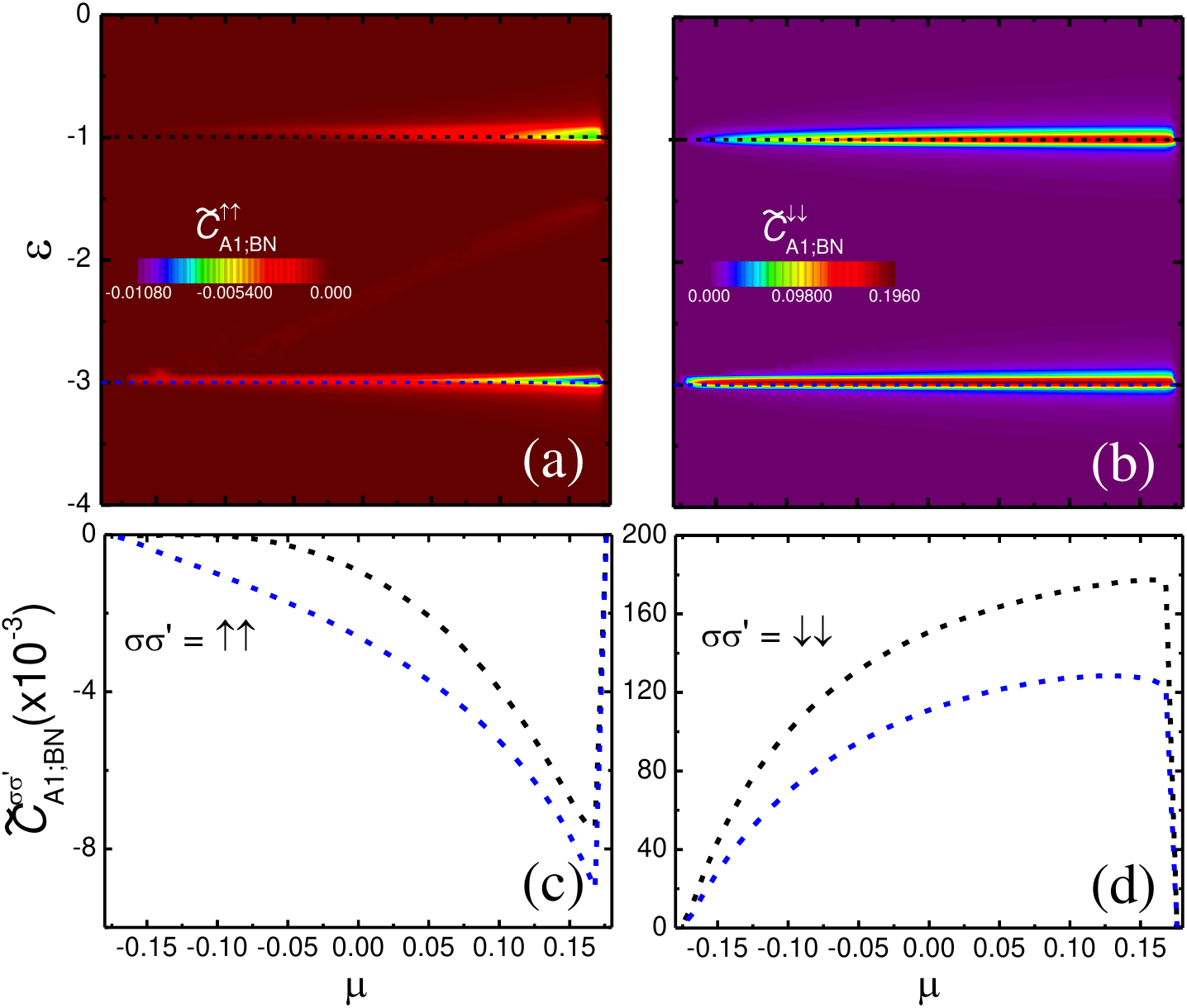}
\caption{Contour plot of $\tilde C_{A1;BN}^{\sigma\sigma^{\prime}}$ versus impurity occupation $\varepsilon$ and chemical potential $\mu$ for the impurity coupled at the chain's edge, for (a) spin $\sigma\sigma^{\prime}=\uparrow\uparrow$ and (b) spin $\sigma\sigma^{\prime}=\downarrow\downarrow$. (c)-(d) are the curves for $\varepsilon= -1.0$ (black dashed line) and $-3.0$ (blue dashed line) for spin $\sigma\sigma^{\prime}=\uparrow\uparrow$ and $\sigma\sigma^{\prime}=\downarrow\downarrow$, respectively. The parameters used in all panels are: $N=162$, $\Delta=0.1$, $\lambda_{\rm SO}=0.1$, $V_{Z}=0.2$, $t_{0}=0.1$ and $U_{I}$=2.0.}
\label{fig5}
\end{figure}

This scenario changes and becomes more interesting when the impurities are populated by decreasing $~\varepsilon$. In Fig.~\ref{fig5}(a) and \ref{fig5}(b), we show the contour plot of the spin resolved leaked Majorana correlation for the impurities located at the chain's edges, as function of impurity level ($\varepsilon_{i=1,2}$) and chemical potential ($\mu$), while fixing the parameters $\lambda_{\rm SO}$, $\Delta$, $V_{Z}$, $t_{0}$ and $U_{I}$. For the system within the topological phase (inside the expected range of $\mu_c$), one can notice two peaks in the leaked Majorana correlation for both spins \emph{up-up} [Fig.~\ref{fig5}(a)] and \emph{down-down} [Fig.~\ref{fig5}(b)], for lower values of  $\varepsilon$. 
The sharp  peaks (horizontal lines) appearing in these figures  signal the the Hubbard energies crossing the zero-energy Majorana modes. Indeed, from Hamiltonian Eq. \eqref{eq:Hamilt2}, one can see that this  occur when $\tilde \varepsilon=\varepsilon+t=0\Rightarrow \varepsilon=-t=-1$ and $\tilde \varepsilon+U_{I}=\varepsilon+t+U_{I}=0\Rightarrow \varepsilon=-t-U_{I}=-3$, which is precisely the positions observed in Figs.~\ref{fig5}(a) and \ref{fig5}(b). There is an additional fainting peak in Fig.~\ref{fig5}(a) whose position depends on the chemical potential $\mu$, this is associated to a spin-resolved population inversion which we will discuss below. 
Figures~\ref{fig5}(c) and \ref{fig5}(d) show $\tilde C_{A1;BN}^{\uparrow\uparrow}$ and $\tilde C_{A1;BN}^{\downarrow\downarrow}$, respectively, for specific values of $\varepsilon$ [along the horizontal dashed lines in panels \ref{fig5}(a) and \ref{fig5}(b)]. Red and blue lines correspond to the correlations along $\varepsilon =-1.0t$ and $-3.0t$ respectively. We can clearly see the monotonic increase as $\mu$ increases, with a relative absolute value around the same value for both spin component \emph{up-up} and \emph{down-down}.   

\begin{figure}[t!]
\includegraphics[scale=0.50]{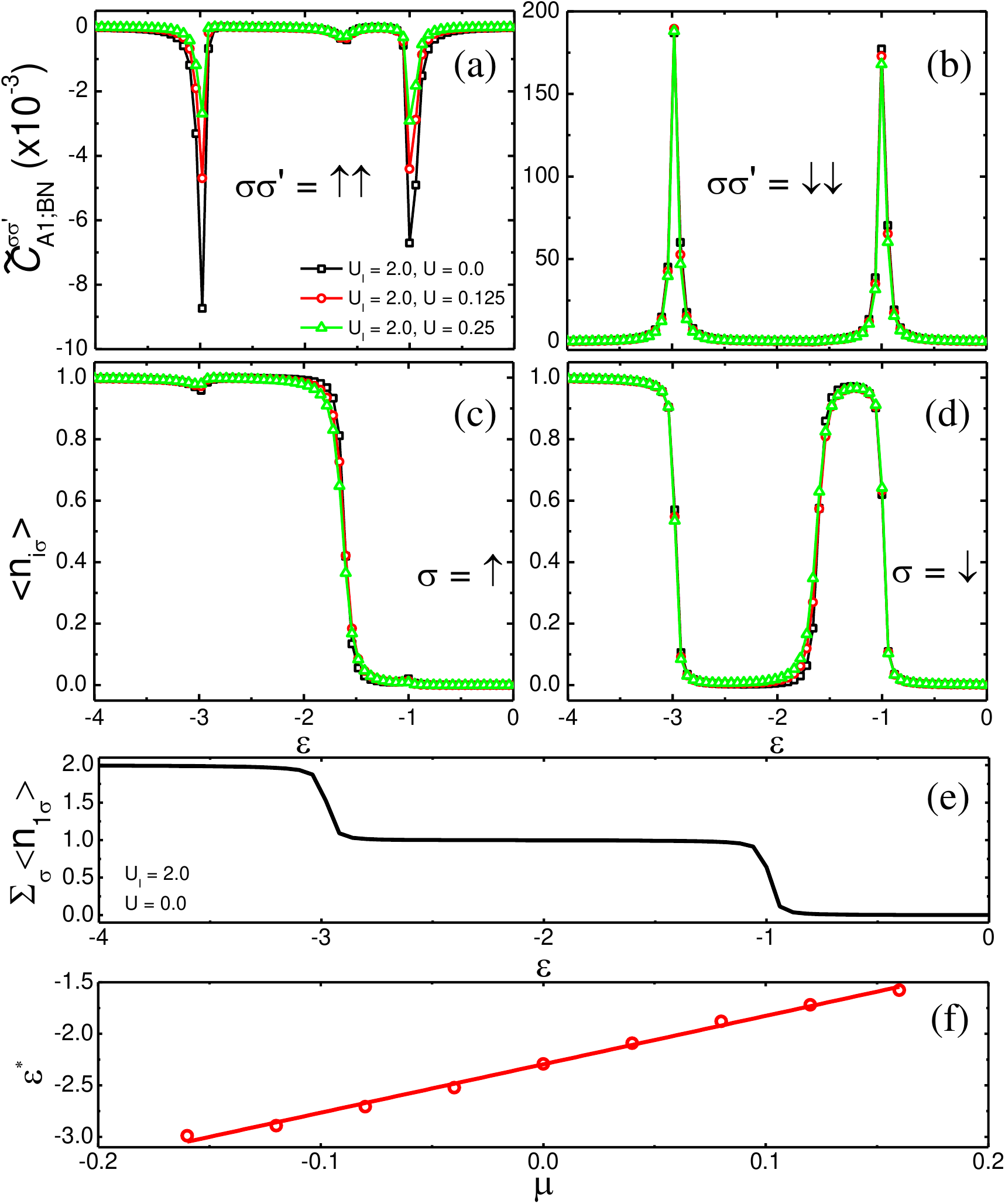}
\caption{$\tilde C_{A1;BN}^{\sigma\sigma^{\prime}}$ versus impurity occupation $\varepsilon$, for different values of $e$-$e$ interaction at the chain and the impurities, for (a) spin $\sigma\sigma^{\prime}=\uparrow\uparrow$ and (b) spin $\sigma\sigma^{\prime}=\downarrow\downarrow$, respectively. (c)-(d) Show the impurity occupation (impurity 1 or N have the same curves) associated to panels (a) and (b). (e) Total electron occupation for the case $U_{\rm I}$=2.0 and $U = 0.0$. (f) Impurity energy level position $\varepsilon^*$ where the population inversion occur  as a function of chemical potential for the same set of parameters used in Fig.~\ref{fig5}.  The parameters used in all panels are: $N = 162 $, $U_{\rm I}=2.0$, $\Delta = 0.1$, $\lambda_{\rm SO} =0.1$, $V_{Z} =0.2$, $t_{0}=0.1$ and $\mu = 0.15$, except in (f) where $\mu$ varies.} 
\label{fig6}
\end{figure}

Now let us analyze what happens in the presence of $e$-$e$ interactions at the impurities ($U_{\rm I}$) and the quantum wire ($U$). To this end, we calculate $\tilde C_{A1; BN}^{\sigma\sigma^{\prime}}$ for a fixed value of chemical potential $\mu=0.15$ and varying $\varepsilon$ . At this value of $\mu$ the system is in the topological regime and the correlation is around its maximum absolute value as seen in Figs.~\ref{fig5}(c) and \ref{fig5}(d). In Figs.~\ref{fig6}(a) and \ref{fig6}(b) we show $\tilde C_{A1;BN}^{\uparrow\uparrow}$ and $\tilde C_{A1; BN}^{\downarrow\downarrow}$, respectively, as function of $\varepsilon$ for $U_{\rm I}=2$ and different values of $U$. Again, as already observed in the previous results, the $\sigma\sigma^{\prime}$=$\downarrow\downarrow$ component has the dominant value, and two high peaks appear separated by the value of $e$-$e$ interaction $U_{\rm I}=2$. We note that the effect of $U$ is more relevant for the spin \emph{up-up} component as it reduces the amplitude of the Coulomb blockade peaks in the correlations as noticed in Fig.~\ref{fig6}(a). For the spin \emph{down-down} component we see that the effect of $U$ is very marginal as we can barely distinguish the peaks in Fig.~\ref{fig6}(b).

As discussed above, the peaks in the correlations are associated to charge fluctuations separated by a Coulomb blockade region. It is therefore useful to analyze the behavior of the electron occupation  $\langle n_{i\sigma} \rangle$  of the impurities as we change their level positions. Since the impurity levels are assumed to be identical $\varepsilon_1=\varepsilon_N=\varepsilon$, we also have $\langle n_{1\sigma}\rangle=\langle n_{N\sigma}\rangle=\langle n_{\sigma}\rangle$.  Figures \ref{fig6}(c) and \ref{fig6}(d) shows the  occupations $\langle n_{\uparrow} \rangle$ and $\langle n_{\downarrow} \rangle$, respectively vs $\varepsilon$ for $U_{\rm I}=2$ and different values of $U$. Like the correlation, the occupation is quite insensitive to the effect of $U$ such that the curves almost collapse on each other. There is, however, a very interesting charge-discharge process for the spin \emph{down} component observed in Fig.~\ref{fig6}(d). As $\varepsilon$ decreases, we see a rapid increase of $\langle n_{\downarrow}\rangle$ at $\varepsilon\approx -1$ while $\langle n_{\uparrow}\rangle$ remains at zero. Then at $\varepsilon \approx -1.6$ $\langle n_{\downarrow}\rangle$ drops to zero and $\langle n_{\uparrow}\rangle$ goes to the unity. The physical origin of this spin-\emph{down} discharging process is quite subtle and can be understood as follows: although both spin components of a given impurity level are coupled to the chain, which produces a broadening in the levels for both spins, it is larger for the spin-\emph{down} component. This is  because of the Zeeman potential necessary for the topological regime, increases the spin-\emph{down} density of states of the chain, suppressing the spin-\emph{up} component. Therefore, as $\varepsilon$ decreases  $\langle n_{\downarrow} \rangle$ start increasing before $\langle n_{\uparrow} \rangle$. Because of Coulomb repulsion $U_{\rm I}$, spin \emph{up} electrons are prevented to hop into the impurity. This picture remains as $\varepsilon$ decreases until the situation reverses and it becomes energetically more favorable for the impurity to host one electron with spin \emph{up}, preventing the presence of the spin \emph{down} electron. This explains the drop of $\langle n_{\downarrow} \rangle$ to zero accompanied by the jump of $\langle n_{\downarrow} \rangle$ to the unity. This new situation remains unchanged until $\varepsilon$ is low enough so that Coulomb repulsion energy is compensated by the presence of a second electron. This occur for $\varepsilon\approx -3$, where $\langle n_{\downarrow} \rangle$ also goes to the unity. Figure \ref{fig6}(e) shows the total occupation $\langle n\rangle=\langle n_{\uparrow} \rangle+\langle n_{\downarrow} \rangle$ that exhibits only two jumps (from zero to one and from one to two), which shows a perfect inversion of \emph{up} and \emph{down} occupations in the impurity.
The energy position  where this occupation inversion takes place depends on the broadening of the levels by the chain. The broadening, in turn, is strongly dependent on the chemical potential $\mu$. Interestingly, the jump in the occupation $\langle n_\uparrow\rangle$ from zero  (accompanied by a drop of $\langle n_\downarrow\rangle$) results in the fainting peak observed in Fig.~\ref{fig5}(a) for the correlation $\tilde C_{A1;BN}^{\uparrow\uparrow}$, whose position $\varepsilon^*$ depends on $\mu$. In Fig.~\ref{fig6}(f) we show $\varepsilon^*$ vs $\mu$, where $\varepsilon^*$ is the value of the impurity levels at which the occupation $\langle n_\downarrow \rangle=\langle n_\uparrow \rangle = 0.5$ when the populations invert [see  Fig.~\ref{fig6}(c) and \ref{fig6}(d)]. Finally, we should mention that this spin-dependent occupation inversion here is very much similar to orbital-dependent occupation inversion predicted by one of us in Refs.~\cite{PhysRevB.83.125404} and \cite{PhysRevB.84.205320}.


%

\section{Conclusion}
\label{conclusions}
In summary, we have studied the presence of Majorana modes in a topological superconducting quantum wire coupled to two quantum impurities. Within the DMRG approach we have calculated impurity-impurity Majorana correlations in the trivial and topological phases of the system. Our results show that when the impurities are coupled directly to the ends of the wire, there is an enhancement of the Majorana correlations associated to those Majorana zero energy states that are bound to the edges of the wire. This phenomenon occurs  both in the interacting and non-interacting regimes.
When Coulomb repulsion is taken into account (in the quantum wire and in the impurities), a rich enthralling physics come out in the correlation function. For instance, an $e$-$e$ interaction in the wire leads to an strong modification in the leaked Majorana correlation for higher values of chemical potential (for $\varepsilon$=0), with the topological phase \emph{window} enlarged, akin to the previously reported work \cite{PhysRevB.84.014503}. More interestingly, in the presence of interaction  $U_{\rm I}$ in the impurities, by  varying their energy levels $\varepsilon$, we observe two pronounced peaks appearing Hubbard energies cross the Majorana zero-energy level. These peaks are robust against $e$-$e$ interaction  $U$ in the quantum wire. Moreover, because of the presence of a Zeeman potential in the wire breaks spin degeneracy in the occupation is broken. This TR asymmetry combined with  a strong Coulomb repulsion  induces a rich spin-dependent population inversion in the impurities occupation.  Our results not only show the richness physics displayed by these systems in the topological regime (where there is a leaked Majorana in the impurities), but also suggests a further possible manner to detect MFs through quantum correlation phenomena, which may be accessible by exploring thermal transport \cite{PhysRevB.89.165416}, inelastic x-ray scattering \cite{TREBST2022} and nonlocal electronic transport experiments \cite{PhysRevB.97.045421,PhysRevLett.124.036801}.

\acknowledgments
EV acknowledges CNPq (process 311366/2021-0) and FAPEMIG (process PPM-00631-17) and GSD thanks the Universidade Federal de Uberl\^andia for hospitality, where this work was conducted.
%

\begin{thebibliography}{47}
\bibitem [{not()}]{note1}%
	\BibitemOpen
	\href@noop {} {}\bibinfo {note} {In the DMRG calculation, we have set a
		number of sweep of 60 with noise to help convergence. A gradually bond
		dimension raising from 10 to 200, cutoff of $10^{-7}$ and a number of
		iteration of 10.}\BibitemShut {Stop}%
\end{thebibliography}
%
%
\end{document}